\documentclass[aps,prd,preprintnumbers,superscriptaddress,nofootinbib,notitlepage,twocolumn]{revtex4-1}
\usepackage[pdftex]{graphicx}
\usepackage{bm,latexsym,amsmath,amssymb,amsfonts,mathrsfs}
\usepackage{color}
\allowdisplaybreaks[1]
\usepackage[pdftex,colorlinks=true,linkcolor=blue,citecolor=cyan]{hyperref}
\newcommand*{\D}{{\rm d}}
\newcommand*{\mpl}{M_{\rm Pl}}

\begin{document}
\title{Nanohertz gravitational waves from NEC violation in the early universe}
\author{Hiroaki~W.~H.~Tahara}
\email[Email: ]{tahara"at"rikkyo.ac.jp}
\affiliation{Department of Physics, Rikkyo University, Toshima, Tokyo 171-8501, Japan
}
\author{Tsutomu~Kobayashi}
\email[Email: ]{tsutomu"at"rikkyo.ac.jp}
\affiliation{Department of Physics, Rikkyo University, Toshima, Tokyo 171-8501, Japan
}
\begin{abstract}
We study nanohertz gravitational waves
relevant to pulsar timing array experiments
from quantum fluctuations in the early universe with null energy condition (NEC) violation.
The NEC violation admits accelerated expansion with the scale factor
$a\propto (-t)^{-p}$ ($p>0$),
which gives the tensor spectral index $n_t=2/(p+1)>0$.
To evade the constraint from Big Bang nucleosynthesis (BBN), 
we connect the NEC-violating phase to a subsequent short
slow-roll inflationary phase which ends with standard reheating,
and thereby reduce the high frequency part of the spectrum.
An explicit model is constructed within the cubic Horndeski theory
which allows for stable violation of the NEC.
We present numerical examples of the background evolution having
the different maximal Hubble parameters (which determine the peak amplitude of
gravitational waves),
the different inflationary Hubble
parameters (which determine the amplitudes of high frequency gravitational waves),
and different durations of the inflationary phase
(which essentially determine the peak frequency of the spectrum).
We display the spectra with $n_t=0.8$, $0.9$, and $0.95$ for $f\lesssim 1/{\rm yr}$,
which are consistent with the recent NANOGrav result. We also 
check that they do not contradict the BBN constraint.
We discuss how the nearly scale-invariant spectrum of curvature perturbations is produced in the NEC-violating phase.
\end{abstract}
\preprint{RUP-20-32}
\maketitle
\section{Introduction}

The stochastic gravitational wave background is
of great interest of cosmologists and astrophysicists
since the detection of its power spectrum reveals to us 
high-energy events and their unknown physics in the early universe.
The power spectrum for $f\lesssim 10^{-15}\,{\rm Hz}$ is constrained 
by experiments of the cosmic microwave background (CMB),
whereas that in the high-frequency range is constrained by 
experiments of pulsar-timing arrays and space/ground-based interferometers.
Recently, the NANOGrav collaboration reported that they found a strong preference 
for a stochastic common-spectrum process in their
12.5-yr data set~\cite{Arzoumanian:2020vkk}
although definitive evidence of quadrupolar interpulsar
correlations~\cite{Hellings:1983fr} is lacked.
If the common-spectrum process is due to supermassive black hole binaries,
it would be the first evidence of their formation and coalescence.
Other possible explanations for the NANOGrav result have been proposed by recent studies~\cite{Ellis:2020ena, Blasi:2020mfx, Buchmuller:2020lbh, Samanta:2020cdk, Nakai:2020oit, Addazi:2020zcj, Ratzinger:2020koh, Neronov:2020qrl, Abe:2020sqb,Vaskonen:2020lbd, DeLuca:2020agl, Kohri:2020qqd, Domenech:2020ers, Bhattacharya:2020lhc, Namba:2020kij, Kitajima:2020rpm, Bian:2020bps, Vagnozzi:2020gtf, Li:2020cjj}
from the view point of
cosmic strings~\cite{Ellis:2020ena, Blasi:2020mfx, Buchmuller:2020lbh, Samanta:2020cdk},
dark phase transition~\cite{Nakai:2020oit, Addazi:2020zcj, Ratzinger:2020koh},
QCD phase transition~\cite{Neronov:2020qrl, Abe:2020sqb}, and
second-order gravitational waves from 
primordial black holes~\cite{Vaskonen:2020lbd, DeLuca:2020agl, Kohri:2020qqd, Domenech:2020ers, Bhattacharya:2020lhc}
or resonant amplification of other fields~\cite{Namba:2020kij,Kitajima:2020rpm}.

Among a variety of possibilities, in this paper we discuss whether it is possible
to generate gravitational waves of inflationary origin~\cite{Guzzetti:2016mkm}
that have amplitudes large enough to be
detected in the nanohertz range with pulsar timing arrays while evading
other observational constraints.
Taking the tensor amplitude and spectral index simply as parameters,
one can derive the region of the parameter space consistent with
the NANOGrav result at $f\sim 1/$yr and the Planck limit at $f\lesssim 10^{-15}\,$Hz.
However, if one assumes that the gravitational wave energy density $\Omega_{\rm GW}(f)$
has a single power-law form with $n_t\sim 1$ up to the high frequency end of the spectrum,
Big Bang nucleosynthesis (BBN) then excludes the entire parameter space~\cite{Vagnozzi:2020gtf}.
To explore the inflationary explanation for the NANOGrav result,
we therefore have to resolve the following issues.
First of all, it is difficult to produce such an extremely blue gravitational wave
spectrum directly from inflation.
A blue inflationary gravitational wave spectrum implies the violation of the
null energy condition (NEC), which usually results in some kinds of instabilities.
Even if stable violation of the NEC were possible,
generating an extremely blue spectrum with $n_t\sim 1$ would be even more difficult
because the Hubble expansion rate that determines the tensor
amplitude is supposed to be nearly constant during inflation.
Furthermore, the spectrum must be less steep for $f\gtrsim 1/$yr
to evade the BBN constraint~\cite{Kuroyanagi:2014nba}.

At the level of the background solution, one can
obtain NEC-violating inflation (super-inflation, $\dot H>0$) with a phantom scalar
field~\cite{Piao:2003ty,Piao:2004tq,Piao:2004jg,Piao:2006jz,Piao:2007sv}
or a k-essence field~\cite{Baldi:2005gk}.
Unfortunately, at the level of linear perturbations
all of these earlier models are unstable. It was, however, pointed out that
stable violation of the NEC is possible in a consistent effective field
theory~\cite{Creminelli:2006xe}. Indeed, galileon and Horndeski
theories~\cite{Horndeski:1974wa,Deffayet:2011gz,Kobayashi:2011nu,Kobayashi:2019hrl} admit
stable NEC-violating inflation~\cite{Deffayet:2010qz,Kobayashi:2010cm}.
(See also~\cite{Anisimov:2005ne} for a higher-derivative scalar-field theory
admitting a similar super-accelerating phase.)
In the same context, a more radical scenario has been proposed
in which the universe starts expanding from Minkowski
(galilean genesis)~\cite{Creminelli:2010ba}.
Variants of the galilean genesis scenario are found in~\cite{Liu:2011ns,Liu:2012ww,Creminelli:2012my,Hinterbichler:2012fr,Hinterbichler:2012yn,Liu:2013xt,Elder:2013gya,Pirtskhalava:2014esa,Nishi:2015pta,Kobayashi:2015gga,Cai:2016gjd,Nishi:2016ljg,Volkova:2019jlj,Ilyas:2020zcb}.
Some of these introduce an NEC-violating phase connected smoothly to
a subsequent quasi-de Sitter inflationary phase~\cite{Pirtskhalava:2014esa,Kobayashi:2015gga}.
See~\cite{Rubakov:2014jja} for a review on NEC violation and
its cosmological consequences.
An NEC-violating inflationary solution has also been considered
in different contexts such as loop quantum gravity~\cite{Copeland:2007qt,Copeland:2008kz}.

In the above NEC-violating examples of the early universe scenarios,
gravitational waves originated from
quantum fluctuations of the metric typically have a blue spectrum.
This is because the amplitude of the metric perturbations is basically given by
$h_{ij}\sim H/\mpl$, where $H$ is the Hubble parameter at the horizon crossing time
(which is increasing when the NEC is violated)
and $\mpl$ is the Planck mass.
Note, however, that its increasing rate, $\dot H/H^2$, is small
in quasi-de Sitter inflation and consequently the tensor spectral index
$n_t$ is much smaller than one. (See, e.g.,~\cite{Mishima:2019vlh,Capurri:2020qgz}
for recent analyses of a blue tensor tilt from slow-roll inflation.)
Therefore, the expansion of the early universe must be accelerating
and significantly away from de Sitter
to produce gravitational waves with $n_t\sim 1$.
(There is a possibility that the primordial tensor spectrum
is blue-tilted due to the time variation of the tensor propagation speed~\cite{Cai:2015yza,Cai:2016ldn},
but its effect can be absorbed entirely into the time variation of
the Hubble parameter in the Einstein frame~\cite{Creminelli:2014wna}.
A non-standard symmetry-breaking
pattern~\cite{Gruzinov:2004ty,Endlich:2012pz,Cannone:2014uqa,Bartolo:2015qvr,Ricciardone:2016lym,Fujita:2018ehq}
and a non-Bunch-Davis initial state~\cite{Ashoorioon:2014nta} can also yield
a blue tensor tilt.)

The purpose of the present paper is to propose an early universe scenario
that can generate large primordial gravitational waves
in the nanohertz range and at the same time is consistent with other observations.
Our scenario is composed of NEC-violating super-inflation
and a subsequent (relatively short)
phase of NEC-preserving slow-roll inflation.
We construct an explicit model within the cubic Horndeski theory.
Our scenario is closely related to, but different in many aspects from,
those in~\cite{Pirtskhalava:2014esa,Kobayashi:2015gga}.

This paper is organized as follows.
In the next section, we give a brief sketch of the scenario.
We then present in more detail the explicit Lagrangian for our model
and study the background cosmological evolution in Sec.~\ref{sec3}.
The gravitational wave spectrum generated in our model is presented in Sec.~\ref{sec4}.
The mechanism for producing scalar perturbations is discussed in Sec.~\ref{sec5}.
Section~\ref{sec6} is devoted to conclusions.

\section{Sketch of the Idea}\label{sec2}

\begin{figure}[tb]
    \begin{center}
            \includegraphics[keepaspectratio=true,height=55mm]{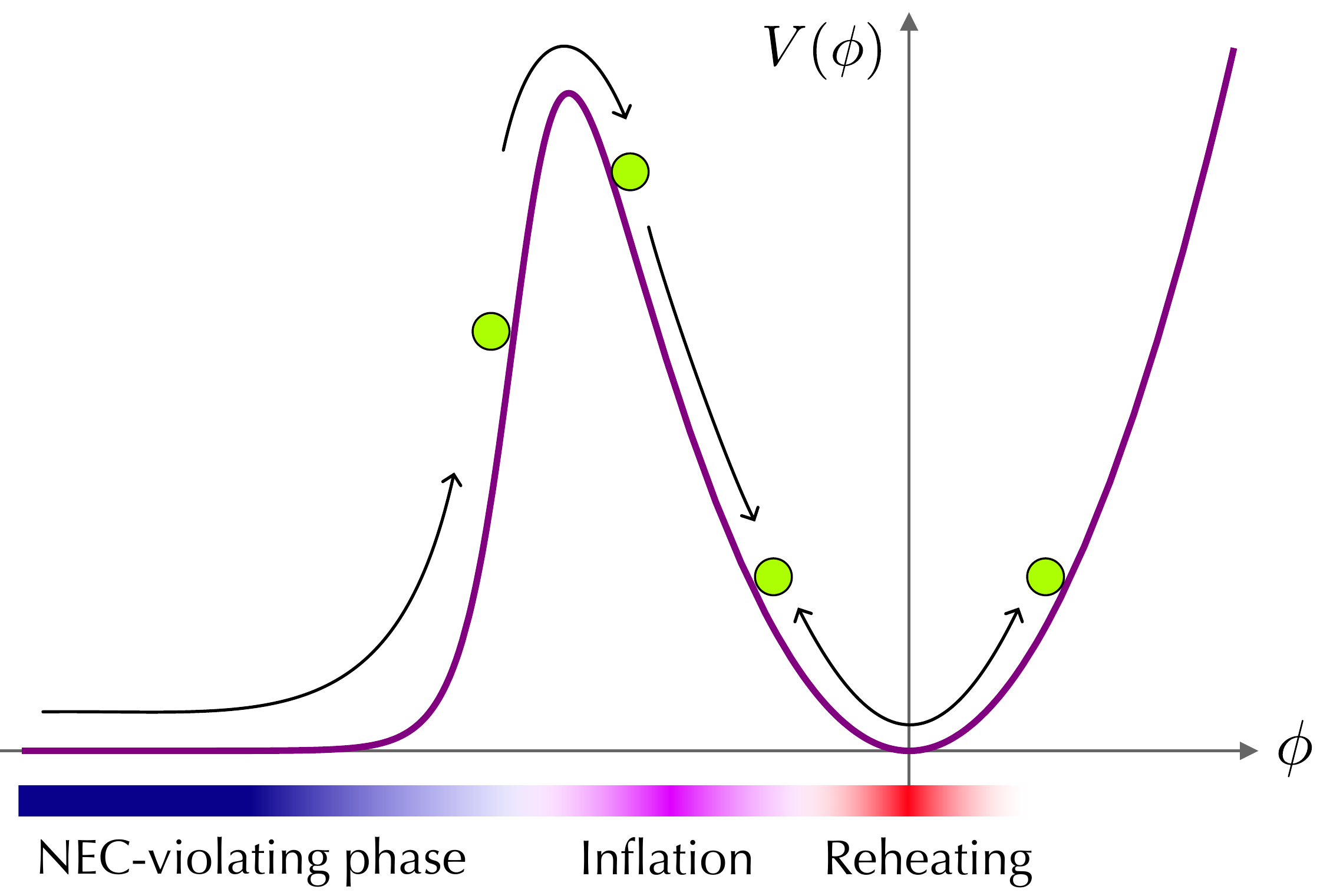}
       \caption{The shape of the potential and the dynamics of the scalar field
       in our scenario.
  	}
       \label{fig:potential.pdf}
  	\end{center}
  \end{figure}

Let us begin with introducing our basic idea
for generating an extremely blue tensor spectrum
in the early universe.
Our scenario is composed of the following three phases:
an NEC-violating phase, subsequent relatively short inflation, and (standard) reheating.
Each phase is characterized by the different dynamics of a scalar field in the potential,
as shown schematically in Fig.~\ref{fig:potential.pdf}.

The scale factor and the Hubble parameter during the NEC-violating phase we consider
are given, respectively, by 
\begin{align}
    a\propto (-t)^{-p},\quad H=\frac{p}{-t},\label{eq:Hubble-NECV}
\end{align}
where $t\,(<0)$ is the proper time and $p\,(>0)$ is a constant.
Clearly, the NEC is violated: $\dot H= p/(-t)^2>0$.
Note that in this phase the universe undergoes an
accelerated expansion, $\ddot a/a=p(1+p)/(-t)^2>0$, so that
the horizon and flatness problems are resolved in the same way
as in standard quasi-de Sitter inflation.
The amplitude of tensor modes is basically determined by
the Hubble parameter at horizon crossing.
In this NEC-violating phase where $H$ is increasing with time as in Eq.~\eqref{eq:Hubble-NECV},
the tensor spectral index is given by 
\begin{align}
    n_t=\frac{2}{p+1}, \label{eq:tensor-index}
\end{align}
and hence $0<n_t<2$ for $p>0$.

The cosmic expansion with $\dot H>0$ is caused by
a scalar field having (effectively) negative kinetic energy
and climbing up the potential.
It is likely that the curvature perturbation exhibits ghost/gradient
instabilities around such a background, which indeed occurs in
$P(\phi,(\partial\phi)^2)$ theory. However, by adding a galileon-type interaction term
$\sim (\partial\phi)^2\Box\phi$ to the Lagrangian,
one can stabilize such an NEC-violating background,
as is demonstrated in Refs.~\cite{Creminelli:2010ba,Deffayet:2010qz,Kobayashi:2010cm}.
We also use this idea to stabilize the NEC-violating phase.

The first phase described above ends when the scalar field goes over
the top of the potential. The Hubble parameter takes the maximum value $H_*$
at that moment.
The scalar field then rolls down the potential and
quasi-de Sitter inflation occurs. This inflationary phase is
rather standard except that its duration is relatively short.
The tensor modes generated during this phase have a nearly flat spectrum.
Finally, the scalar field starts to oscillate around the bottom of the potential
and thereby the reheating process proceeds in a standard way.

In the next section we present a concrete model realizing the entire
cosmological evolution of this scenario.

\section{Model}\label{sec3}

In this section, we seek for the Lagrangian that admits
the aforementioned early universe scenario within
the following subclass of the Horndeski theory:
\begin{align}
    {\cal L}=G_2(\phi,X)-G_3(\phi,X)\Box\phi+\frac{\mpl^2}{2}R,\label{cubic-Horndeski}
\end{align}
where $X=-g^{\mu\nu}\partial_\mu\phi\partial_\nu\phi/2$.
The general background and perturbation equations in this theory
are summarized in the appendix.

\subsection{The Lagrangian for the NEC-violating phase}\label{subsec3a}

Before investigating the full Lagrangian
that is capable of describing the entire cosmological evolution
sketched in the previous section,
let us focus on the Lagrangian just for the NEC-violating phase.
The Lagrangian we consider for the moment is given by
\begin{align}
{\cal L}=\frac{\mpl^2}{2}R +\beta X-V(\phi)-\frac{\alpha \mpl}{U(\phi)}X\Box\phi,
\label{Lag:NEC-V}
\end{align}
where 
\begin{align}
   V(\phi)=U(\phi)=U_0e^{2\phi/\mpl},
   \quad U_0={\rm const}>0,
   \label{Pot:U=V}
\end{align}
and $\alpha$ and $\beta$ are dimensionless parameters.
In the Horndeski language, this is obtained by taking
$G_2=\beta X-V(\phi)$ and $G_3=\alpha\mpl X/U(\phi)$ in Eq.~\eqref{cubic-Horndeski}.

\begin{figure}[tb]
    \begin{center}
            \includegraphics[keepaspectratio=true,height=95mm]{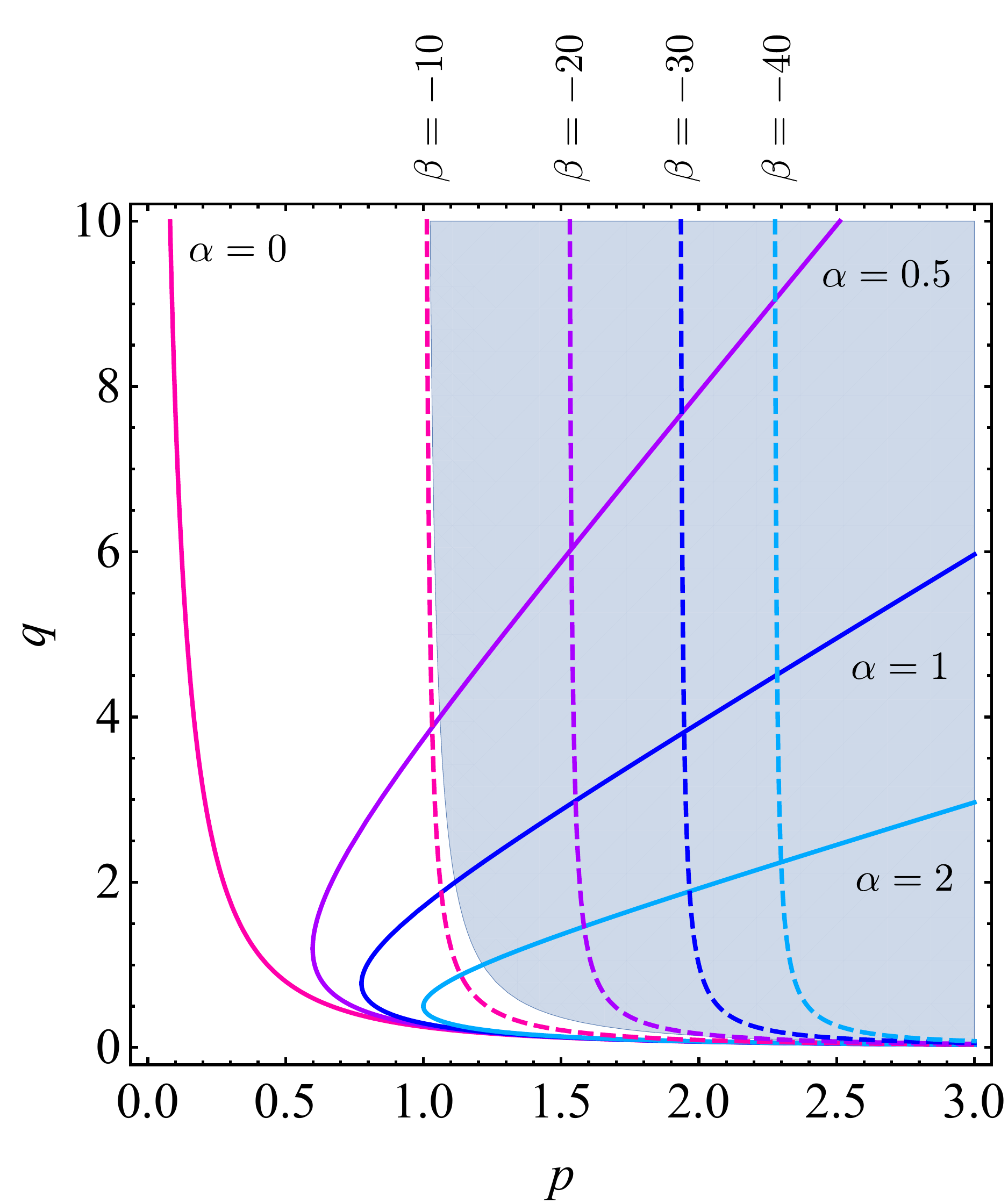}
       \caption{Constant $\alpha$ curves (solid lines) and constant $\beta$ curves
       (dashed lines) in the $p$-$q$ plane. The two stability conditions are
       satisfied in the shaded region.
  	}
       \label{fig:stable.pdf}
    \end{center}
  \end{figure}

It is straightforward to see
from the field equations~\eqref{app:Fridmann}--\eqref{app:Scalar}
that the Lagrangian~\eqref{Lag:NEC-V}
with~\eqref{Pot:U=V} admits the desired NEC-violating solution of the form
\begin{align}
H=\frac{p}{-t},\quad \phi = \mpl \ln \left[\frac{\mpl}{\sqrt{q U_0}(-t)}\right],
\label{Soln:NEC-V}
\end{align}
where $p>0$ and $q>0$ are dimensionless parameters characterizing the solution.
These parameters are determined solely from $\alpha$ and $\beta$:
$p=p(\alpha,\beta)$, $q=q(\alpha,\beta)$.
Compact expressions are obtained if one instead writes $\alpha$ and $\beta$
in terms of $p$ and $q$:
\begin{align}
\alpha = \frac{2p}{q}-\frac{2}{q^2(1+3p)},
\quad
\beta =-2p(2+3p)+\frac{2}{q}.\label{Eq:abpq}
\end{align}
Figure~\ref{fig:stable.pdf} shows constant $\alpha$ and constant $\beta$ curves
in the $p$-$q$ plane. For given $(\alpha,\beta)$, one finds two
sets of $(p,q)$ satisfying $p,q>0$.

In order for the NEC-violating background~\eqref{Soln:NEC-V} to be stable,
it is required that
\begin{align}
{\cal F}_S&=-1+(-1-2p+3p^2)q>0,\label{Stab:1}
\\
{\cal G}_S&=2 p (1+3 p)^3 q^2-3 (1+3 p)^2 q+3>0.\label{Stab:2}
\end{align}
The two conditions are satisfied
in the shaded region in Fig.~\ref{fig:stable.pdf}.
A stable NEC-violating solution can be found for $\alpha>0$ and $\beta<-10$.
Note that all solutions with $p\le  1$ are unstable.
As we have mentioned above, we have two sets of $(p,q)$ for given $(\alpha,\beta)$.
It turns out that at least one of them leads to an unstable background.
(Since this unstable branch typically has $p\gg 1$ and $q\ll 1$,
it is difficult to judge the stability by seeing Fig.~\ref{fig:stable.pdf}.)

\subsection{The full Lagrangian}\label{subsec3b}

We have thus seen that the Lagrangian~\eqref{Lag:NEC-V} with~\eqref{Pot:U=V}
admits the NEC-violating solution~\eqref{Soln:NEC-V},
and there is a parameter region where the solution is stable
against scalar perturbations.
In this subsection, we modify the Lagrangian~\eqref{Lag:NEC-V}
so that it can be used to describe the subsequent inflation and reheating stages
as well as the NEC-violating phase.

The full Lagrangian is given by
\begin{align}
    {\cal L}=\frac{\mpl^2}{2}R+B(\phi)X-V(\phi)-\frac{\alpha\mpl}{U(\phi)}X\Box\phi,
    \label{Lag:full}
\end{align}
where we now take
\begin{align}
    V(\phi)&=\frac{U(\phi)\cdot m^2\phi^2}{2U(\phi)+m^2\phi^2},\label{Pot:full}
    \\
    B(\phi)&=\frac{1+\beta}{2}+\frac{1-\beta}{2}\tanh[c(\phi-\phi_*)/\mpl],\label{Func:kinatic}
\end{align}
with $U(\phi)=U_0e^{2\phi/\mpl}$.
The potential $V(\phi)$ is taken so that $V\simeq U$ for negatively large $\phi$
and $V\simeq m^2\phi^2/2$ for small $\phi$, as shown in Fig.~\ref{fig:potential.pdf}.
Any potential with a similar form is possible as well, but we
use this particular form for simplicity.
The potential peaks at $\phi=\phi_*\,(<0)$, where
\begin{align}
    \phi_*=-\frac{3}{2}\mpl W_0(z),
    \quad 
    z=\frac{2^{4/3}}{3}\left(\frac{U_0}{\mpl^2m^2}\right)^{1/3},
    \label{Def:phi-ast}
\end{align}
with $W_0$ being the Lambert $W$ function (the principal branch of the inverse functions of $z=We^{W}$).
The coefficient of the kinetic term, $B(\phi)$, is taken so that 
$B\simeq \beta \,(<0)$ for $\phi<\phi_*$ and $B\simeq 1$ for $\phi>\phi_*$.
Here, $c$ is some constant of ${\cal O}(1)$--${\cal O}(10)$
whose explicit value is not important.
The sign flip of the kinetic term soon after the end of the NEC-violating phase
is necessary for the subsequent inflation and reheating stages to be
free from ghost instability. This idea is analogous to what is supposed to happen
at the end of k-inflation~\cite{ArmendarizPicon:1999rj}.

The Lagrangian~\eqref{Lag:full} reduces to
Eq.~\eqref{Lag:NEC-V} with~\eqref{Pot:U=V} for negatively large $\phi$,
giving rise to the NEC-violating solution~\eqref{Soln:NEC-V}.
The sign of the kinetic term is flipped when $\phi$ goes over the top of the potential,
and inflation then starts
in the potential $V\simeq m^2\phi^2/2$
provided that $|\phi_*|\gtrsim \mpl$. After inflation,
reheating occurs in a usual way at the bottom of the $\phi^2$ potential.
We thus expect that the scenario outlined in the previous section is
realized by the Lagrangian~\eqref{Lag:full}.
This will be demonstrated explicitly by numerical calculations
in the next subsection.

We have four key quantities characterizing our model:
$p$ in the expansion law in the NEC-violating phase,
the maximum value of the Hubble parameter reached at the end of
the NEC-violating phase (denoted by $H_*$),
the inflationary Hubble scale (hereafter denoted by $H_{\rm inf}$),
and the duration of inflation (hereafter denoted by $\Delta{\cal N}_{\rm inf}$).
These quantities are determined by the four parameters in the Lagrangian~\eqref{Lag:full}
($\alpha$, $\beta$, $U_0$, and $m$).\footnote{The parameter $c$
plays essentially no role in the dynamics.}
As we have already seen, $p$ is determined by $\alpha$ and $\beta$
through the relation~\eqref{Eq:abpq}.
Assuming that the transition from the NEC-violating phase
to inflation is sudden, one can estimate as
\begin{align}
    H_*&\sim \frac{p\sqrt{q U_0}}{\mpl}e^{\phi_*/\mpl}=p\sqrt{\frac{q}{2}}m
    \left(\frac{|\phi_*|}{\mpl}\right)^{3/2},
    \label{Estimate:H-ast}
    \\
    H_{\rm inf}&\sim\frac{m|\phi_*|}{\sqrt{6}\mpl},
    \label{Estimate:H-inf}
\end{align}
where $\phi_*=\phi_*(U_0,m)$ was already defined in Eq.~\eqref{Def:phi-ast}.
In deriving the estimate~\eqref{Estimate:H-ast}, we used
$H_*\sim p/(-t_*)$ and $\phi_*\sim \mpl \ln[\mpl/\sqrt{qU_0}(-t_*)]$,
where $t_*$ is the time at the end of the NEC-violating phase.
Finally, a simple textbook calculation yields
\begin{align}
    \Delta{\cal N}_{\rm inf}\sim \frac{1}{4}\left(\frac{|\phi_*|}{\mpl}\right)^{2}.
    \label{Estimate:N-inf}
\end{align}

As seen from Eq.~\eqref{Estimate:N-inf},
we can adjust the number of
e-folds of inflation by shifting the top of the potential $\phi_*$.
To demonstrate how $\phi_*$ depends on $U_0$, 
it is convenient to introduce a new parameter $\phi_0$ defined as
$\phi_0:=(1/2)\mpl\ln(U_0/\mpl^4)$
(i.e., $U_0 = \mpl^4 e^{2 \phi_0 / \mpl}$).
The shift $\phi_0\to\phi_0+c$ with $c>0$ ($c<0$)
causes left (right) shifting of $U(\phi)$.
The Lambert $W$ function $W_0(z)$ can be approximated roughly by
$W_0(z) \approx \ln{(z/\ln{z})}$ or more crudely $W_0(z) \sim \ln{z}$ for large $z>0$.
Using the latter approximation here in Eq.~\eqref{Def:phi-ast}, we get 
$\phi_* \sim -\phi_0 + {\rm const.}$
We typically take $\phi_0/\mpl \sim \mathcal{O}(10)$ to sustain 
a sufficient duration of inflation.
One might suspect that such a value of $\phi_0$ yielded extremely large $U_0$ 
and the universe experienced super-Planck scale $H$. However,
since $\mpl^2H^2\lesssim V(\phi_*) = (1/2) m^2 \phi_*^3 / (\phi_*-\mpl)$,
it is not the case as long as $\left| \phi_* \right| \ll \mpl^2/m$.
Hereafter we use $\phi_0$ instead of $U_0$.

\subsection{Numerical results}

  \begin{figure}[tb]
    \begin{center}
            \includegraphics[keepaspectratio=true,height=45mm]{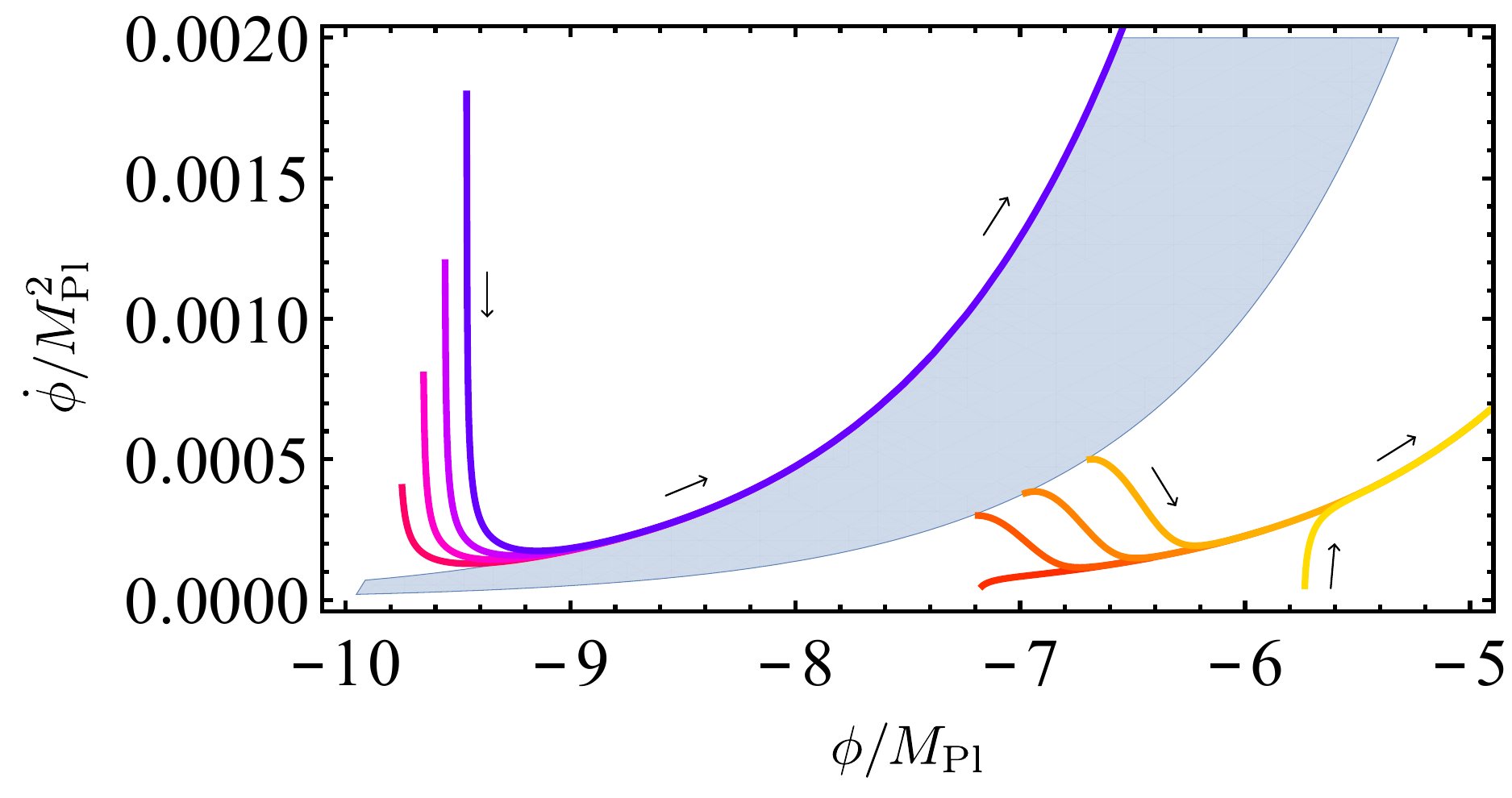}
       \caption{Typical phase-space trajectories. The Friedmann constraint
       does not have real roots for $H$ in the shaded region.
  	}
       \label{fig:attractor.pdf}
    \end{center}
  \end{figure}

Let us first affirm that the NEC-violating solution~\eqref{Soln:NEC-V}
is an attractor. Figure~\ref{fig:attractor.pdf} shows the behavior
of the solutions for $(\alpha,\beta)= (1.09,-12.4)$ with different initial conditions,
obtained by solving the dynamical equations~\eqref{app:Evolution}
and~\eqref{app:Scalar} numerically.\footnote{The initial value for $H$ is determined from
the Friedmann constraint~\eqref{app:Fridmann}. We check that the constraint~\eqref{app:Fridmann}
is satisfied at each time step.}
We see that the trajectories converge to either of the two curves,
$\dot\phi\propto \sqrt{q}e^{\phi/\mpl}$, with different $q$.
In the upper left region the solutions converge to
the stable NEC-violating solution with $(p,q)\simeq (1.2,2)$, while in the lower right region
the solutions converge to the unstable background with $(p,q)\simeq (6.1,8.4\times 10^{-3})$.
The two regions of initial conditions are separated by the ``no-go'' region
(the shaded region in Fig.~\ref{fig:attractor.pdf})
where the Friedmann constraint equation does not have real roots for $H$.
In the following analysis, we assume that the initial conditions for $(\phi,\dot\phi)$
are in the upper left region so that they lead to a stable NEC-violating universe.

\begin{figure}[tb]
    \begin{center}
        \includegraphics[keepaspectratio=true,height=100mm]{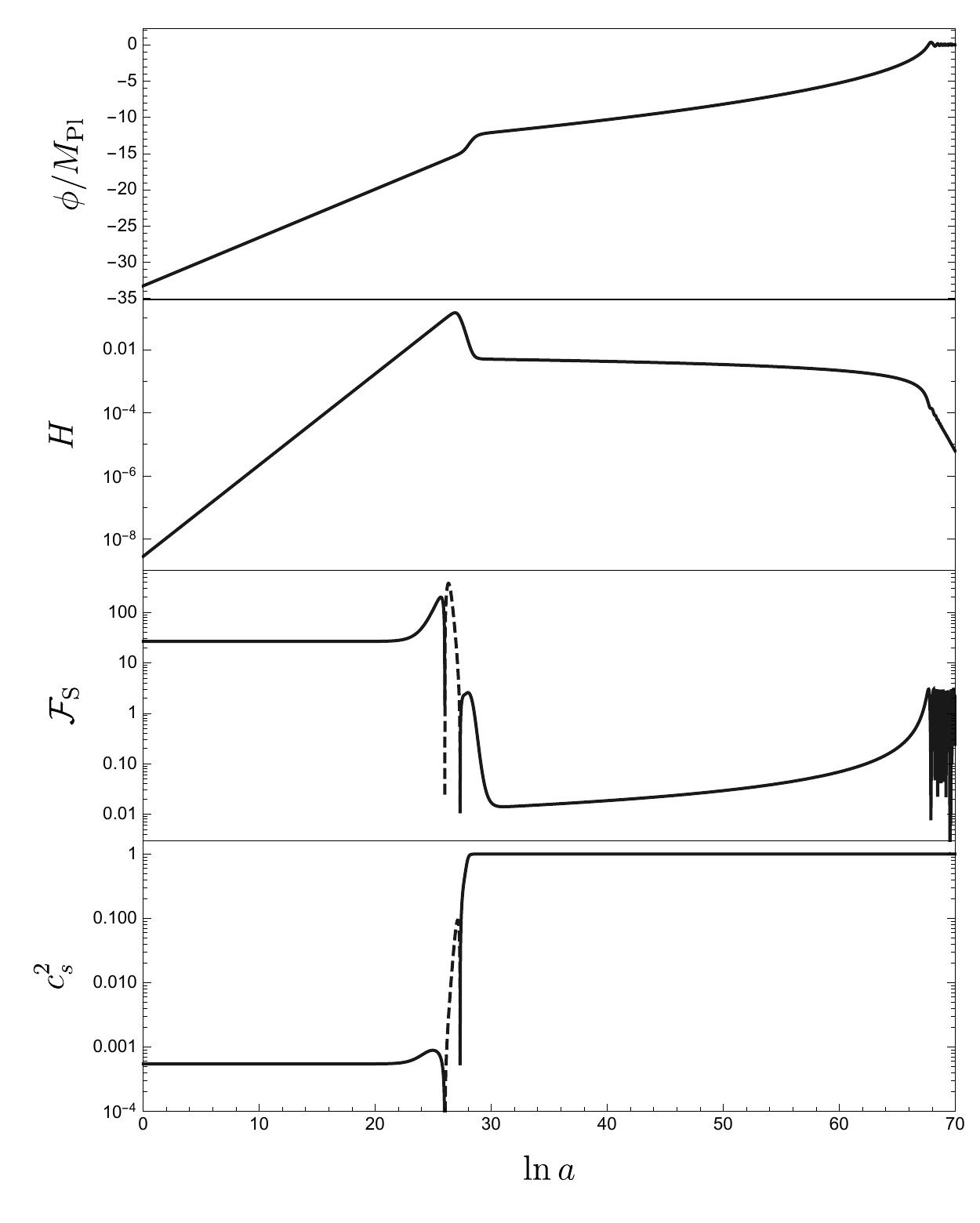}
       \caption{
       Time evolution of $\phi$, $H$, $\mathcal{F}_S$, and $c_s^2 (= \mathcal{F}_S/ \mathcal{G}_S)$.
       The model parameters are chosen as $p=3/2$ ($n_t=0.8$), $q=10$, $m^2=10^{-6}\mpl^2$, 
       and $\phi_0=12\mpl$.
       In the plots, 
       the dashed lines denote the negative value having the same absolute value.
  	}
       \label{fig:time-evolution-of-vars}
    \end{center}
  \end{figure}

We then investigate the entire evolution numerically.
In Figure~\ref{fig:time-evolution-of-vars},
we plot the time evolution 
of $\phi$, $H$, $\mathcal{F}_S$, and $c_s^2 := \mathcal{F}_S/ \mathcal{G}_S$
for the parameter choice
$p=3/2$ ($n_t=0.8$), $q=10$, $m^2=10^{-6}\mpl^2$, and $\phi_0=12\mpl$.
In the NEC-violating phase ($0<\ln a < 26.0$), we have positive $\mathcal{F}_S$.
Note that $H_*$ is as high as $\mathcal{O}(0.1)\times \mpl$.
In the short decelerating phase ($26.0 < \ln a < 27.3$),
$\mathcal{F}_S$ becomes negative,
which does not contradict the stability condition~\eqref{Stab:1} 
since the system is no longer on the attractor~\eqref{Soln:NEC-V}.
This point will be discussed shortly.
In the inflationary phase ($27.3 < \ln a \lesssim 67$), $\mathcal{F}_S$ is approximately given by 
$- \D  \ln H / \D  \ln a$, which is positive in an NEC-preserving universe.
At $\ln a =67.7$, $\phi$ passes the origin and starts to oscillate, 
leading to the expansion of a matter-dominated universe.

Our numerical result implies a potential drawback of the present model
that it exhibits a gradient instability in the short intermediate stage
between the NEC-violating and inflationary phases.
This is similar to what happens at the transition from
the galilean genesis phase to inflation~\cite{Pirtskhalava:2014esa},
and in fact the appearance of gradient instabilities is generic to
non-singular cosmological solutions in galileon/Horndeski
theories~\cite{Libanov:2016kfc,Kobayashi:2016xpl}.
In the present case, however, the instability could in principle be
avoidable by more elaborate choices of the Lagrangian and model parameters,
because the background~\eqref{eq:Hubble-NECV} is
singular in the $t\to-\infty$ limit\footnote{{Since
$\int^t_{-\infty}a\D t<\infty$ for $p>1$, this NEC-violating universe
is past incomplete. In this sense, the present model is
as bad as inflation.}}
 and hence the ``no-go''
theorems in~\cite{Libanov:2016kfc,Kobayashi:2016xpl} do not apply.
The instability issue can also be resolved
by introducing ``beyond-Horndeski'' or higher-order
operators~\cite{Creminelli:2016zwa,Cai:2016thi,Cai:2017tku,Kobayashi:2015gga}.

\section{Gravitational Wave Spectrum}\label{sec4}

Since we do not consider the scalar field derivatively coupled to
the curvature through the so-called
Horndeski $G_4(\phi,X)$ and $G_5(\phi, X)$ terms,
the tensor perturbations $h_{ij}$ evolve following the
same equation as in general relativity.
Their equation of motion is 
\begin{align}
    h_{ij}'' + 2 (a'/a) h_{ij}'-\partial^2 h_{ij} = 0,
\end{align}
where a prime denotes the derivative with respect to
the conformal time $\eta$.
We move to Fourier space, expand $h_{ij}$ into the basis
of the polarization tensors, and write
\begin{align}
    h_{ij}(\eta,\Vec{x})=\frac{1}{(2\pi)^3}\sum_{A=+,\times}\int\D^3k \,
    \tilde h_A(\eta,\Vec{k})e^A_{ij}(\Vec{k})e^{i\Vec{k}\cdot\Vec{x}},
\end{align}
where $e^A_{ij}$ is transverse, traceless, and normalized as $e^A_{ij}e^{A'}_{ij}=\delta^{A}_{A'}$.
Following the usual quantization procedure,
we obtain the canonically normalized positive frequency mode
in the NEC-violating phase as
\begin{align}
    \tilde h_A=\frac{\sqrt{\pi}}{ a \mpl } \sqrt{-\eta} H^{(1)}_{\nu} (-k\eta),
    \label{posfreqmode}
\end{align}
where $H^{(1)}_{\nu}$ is the Hankel function and $\nu := (3p+1)/2(p+1)$.
The initial conditions for the tensor perturbations
are set by Eq.~\eqref{posfreqmode}.
In the superhorizon regime, $|k\eta|\ll 1$, we have
\begin{align}
    k^3|\tilde h_A|^2\simeq \frac{2^{2\nu}\Gamma^2(\nu)}{\pi \mpl^2}
    \left(\frac{2H_c}{2\nu-1}\right)^{2\nu-1}k^{3-2\nu}=
    {\rm const},
\end{align}
where we introduced the constant
$H_c:=Ha^{-1/p}$.
From this, we get the primordial power spectrum of gravitational waves as
\begin{align}
    \mathcal{P}_{{\rm GW}} 
    =2\cdot \frac{k^3}{2\pi^2}|\tilde h_A|^2
                    = \frac{2^{2\nu} \Gamma(\nu)^2}{\pi^3 \mpl^2} 
                    \! \left( \frac{2 H_c}{2\nu-1} \right)^{\! 2\nu-1} \!\! k^{3-2\nu} ,
                    \label{pgw}
\end{align}
Note that for $\nu=3/2$
this reproduces the usual primordial power spectrum in de Sitter inflation.
From Eq.~\eqref{pgw} we see that
the modes which exit the horizon in the NEC-violating phase
have a primordial spectrum with a blue spectral tilt~\eqref{eq:tensor-index},
whereas those which exit the horizon in the subsequent inflationary phase have 
an almost flat primordial spectrum with $n_t =-2\epsilon\, (<0)$,
where $\epsilon$ is the usual small slow-roll parameter,
$\epsilon:=-\dot H/H^2$.
After re-entering the horizon their amplitudes decay proportionally to $1/a$.

\begin{figure}[tb]
    \begin{center}
            \includegraphics[keepaspectratio=true,height=55mm]{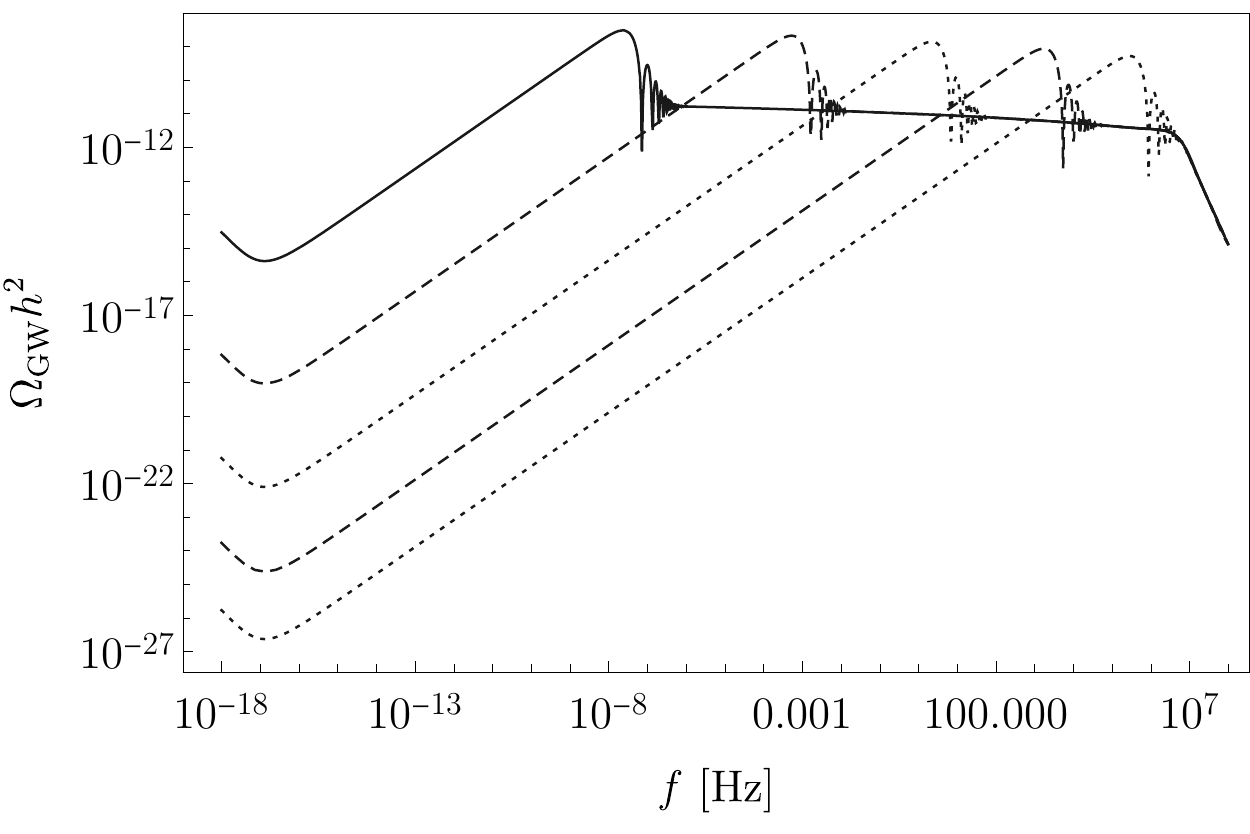}
       \caption{Gravitational wave spectra in the cases of 
       $\phi_0/\mpl=12$ (solid), $10$ (left dashed), $8$ (left dotted), 
       $6$ (right dashed), and $4$ (right dotted).
       The other parameters are the same as in Fig.~\ref{fig:time-evolution-of-vars}. 
  	}
       \label{fig:Energy_phi0}
  	\end{center}
  \end{figure}

The present energy density spectrum of gravitational waves is given by 
\begin{align}
    \Omega_{\rm GW}(k) =\frac{k^5 \left(
    |\tilde{h}_{+} |^2 + |\tilde{h}_{\times} |^2\right) }{24 \pi^2 a^2 H_0^2},
    \label{Def:GWspectrum}
\end{align}
and below we show $\Omega_{\rm GW}(f)$ plots, where $k$ is the comoving wavenumber,
$f$ is the physical frequency $f = k/2\pi a$,
and $H_0$ is the Hubble constant.
Each spectrum in our model is typically composed by the three parts; 
an extremely blue-tilted spectrum in the low frequency range;
{wiggles} in the middle frequency range
(due to the intermediate decelerating stage);
and a slightly red-tilted spectrum in the high frequency range.
We plot the energy density spectra
for different values of $\phi_0$ in Fig.~\ref{fig:Energy_phi0}.
A larger value of $\phi_0$ leads to a larger number of e-folds, a lower peak frequency, 
and as a result larger $\Omega_{\rm GW}(f)$ on CMB scales.
Note in passing that here and hereafter we assume that
the Hubble parameter at which the reheating process becomes so efficient
that the energy is converted from matter to radiation rapidly, $H_{\rm R}$,
is given by $H_{\rm R}=10^{-8}\mpl$.

\begin{figure}[tb]
    \begin{center}
        \includegraphics[keepaspectratio=true,height=55mm]{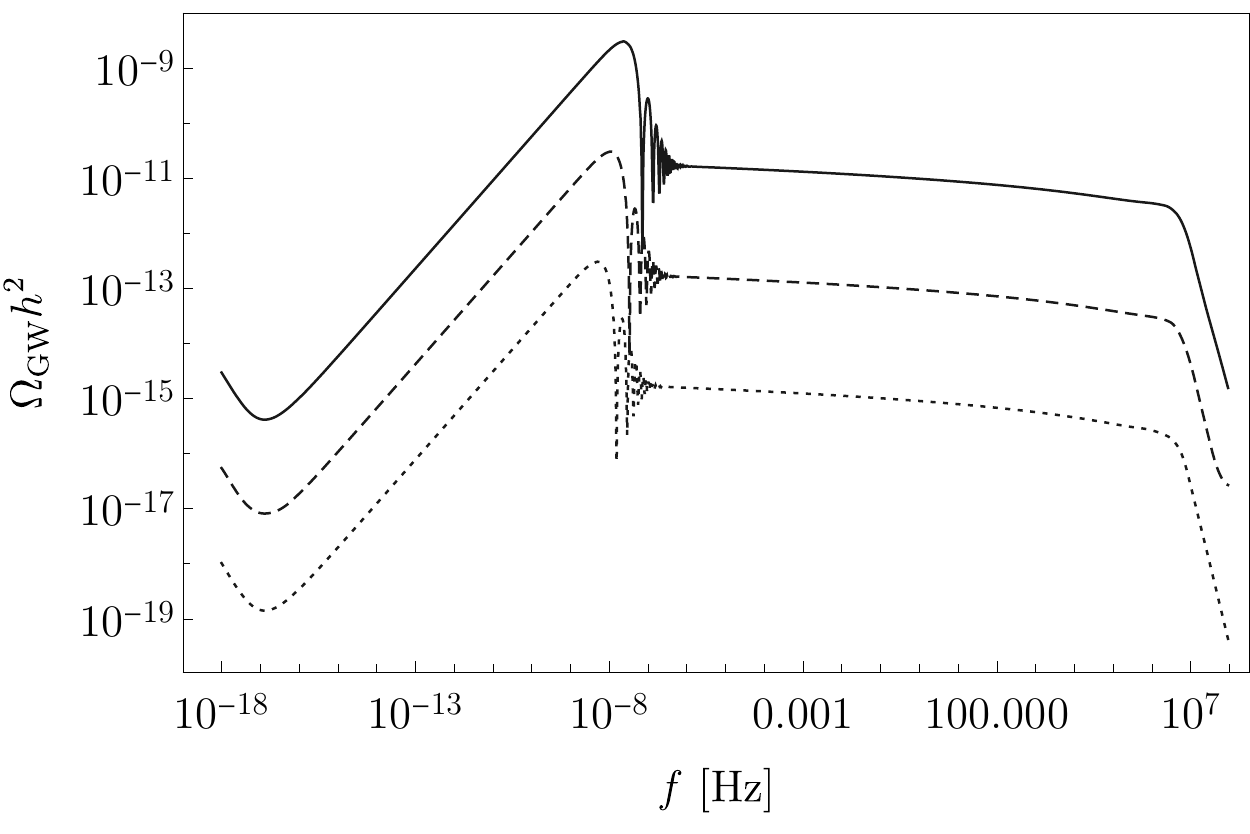}
       \caption{Gravitational wave spectra in the cases of 
       $m^2/\mpl^2=10^{-6}$ (solid), $10^{-8}$ (dashed), and $10^{-10}$ (dotted).
       In each spectrum, we set $\phi_0/\mpl = 12+\ln{(10^3 m/\mpl)}$ to have nearly constant $\phi_*$.
       The other parameters are the same as in Fig.~\ref{fig:time-evolution-of-vars}.
  	}
       \label{fig:Energy_m}
    \end{center}
  \end{figure}

\begin{figure}[tb]
    \begin{center}
        \includegraphics[keepaspectratio=true,height=55mm]{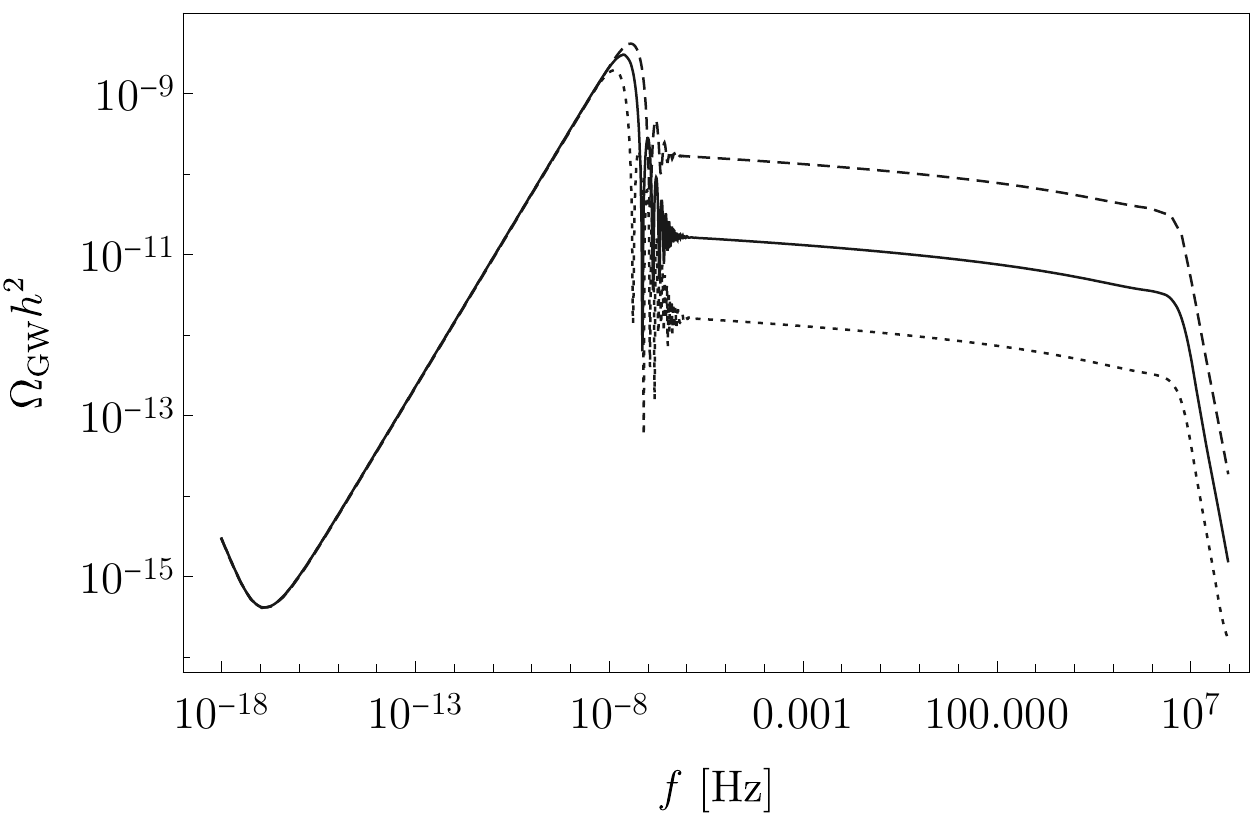}
       \caption{Gravitational wave spectra in the cases of 
       $m^2=10^{-5}$ (dashed), $10^{-6}$ (solid), and $10^{-7}$ (dotted).
       The parameter $q$ are taken as $q m^2=10^{-5}$ 
       in order to produce similar blue-tilted spectra.
       The other parameters are the same as in Fig.~\ref{fig:time-evolution-of-vars}.
  	}
       \label{fig:Energy_qm}
    \end{center}
  \end{figure}

Let us study the effects of changing
the other parameters $m$, $q$, and $p$.
Since we have $|\phi_*|/\mpl \sim \mathcal{O}(1-10)$,
$m$ is the only parameter which determines 
the order of an energy scale of the slow-rolling phase,
although $\phi_*$ also has an implicit logarithmic dependence on $m$ and $\phi_0$.
We plot the energy density spectra for different $m$ in Fig.~\ref{fig:Energy_m}.
From Eq.~\eqref{Estimate:H-ast}, once we set $qm^2$ constant,
we get similar values of $H_*$ even with different $m$.
We show energy density spectra in the case of different $m$ but the same $qm^2$
in Fig.~\ref{fig:Energy_qm}.

\begin{figure*}[tb]
    \begin{center}
            \includegraphics[keepaspectratio=true,height=100mm]{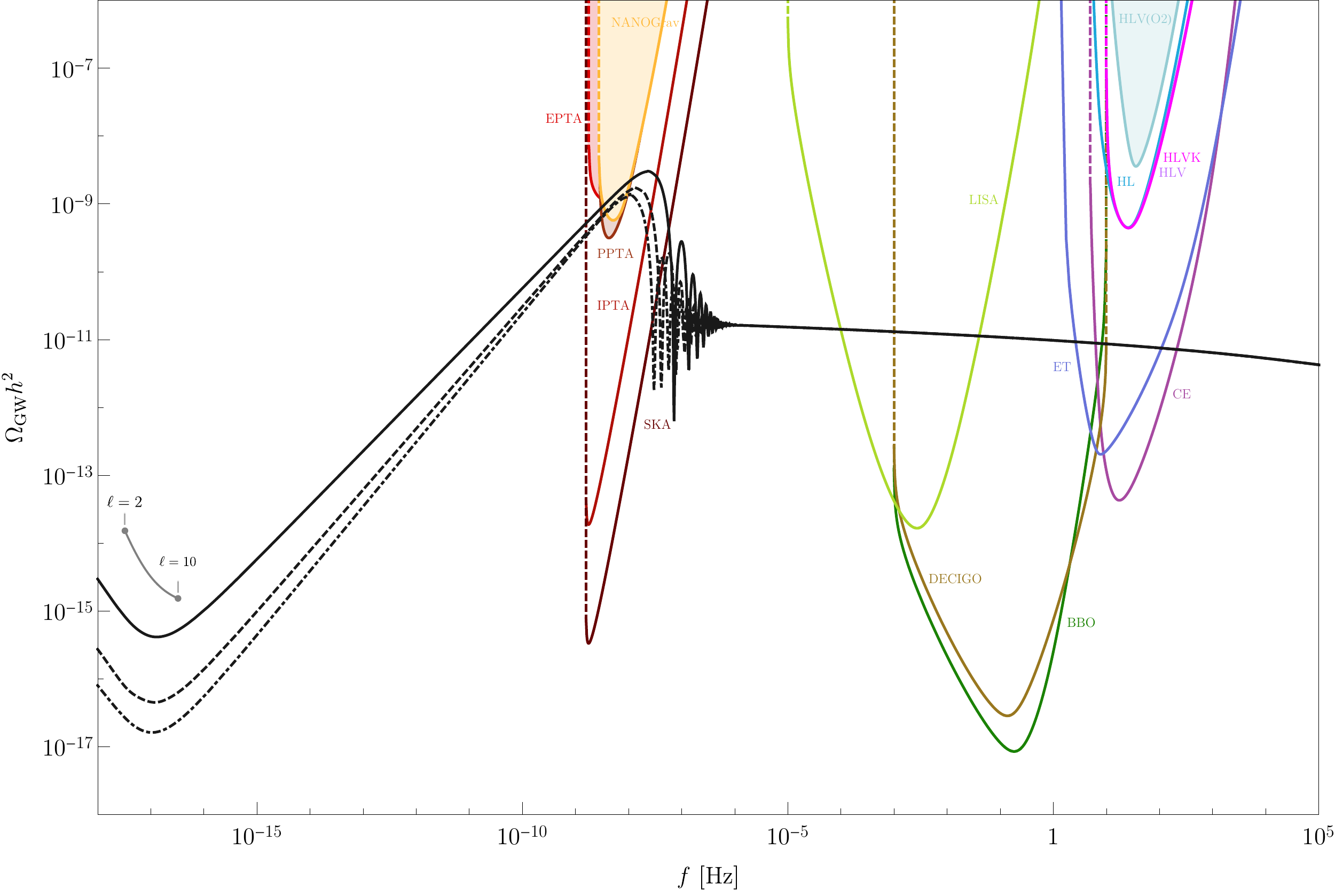}
       \caption{Energy density spectra of gravitational waves in the cases of $n_t=0.8$ (solid), $0.9$ (dashed), and $0.95$ (dot-dashed), 
       which correspond to $p=3/2$, $11/9$, and $21/19$, respectively.
       The other parameters are the same as in Fig.~\ref{fig:time-evolution-of-vars}.
       We give the power-law-integrated sensitivity curves~\cite{Schmitz:2020syl} for present and future experiments of pulsar-timing arrays
       and space/ground-based interferometers.
       For comparison with the lower-frequency constraint, we also plot as the gray line 
       the spectrum with $n_t=0$ and $r=0.07$,
       the upper limit from the {\it BICEP2/Keck Array} and {\it Planck} experiments~\cite{Ade:2018gkx},
       between the frequencies corresponding to multipoles $\ell=2$ and $10$.
  	}
       \label{fig:Energy_nt95}
    \end{center}
  \end{figure*}

To compare our results with the NANOGrav observation~\cite{Arzoumanian:2020vkk},
we parametrize the characteristic strain of gravitational waves 
around the reference frequency $f_{\rm yr}=1/{\rm yr}$ as
\begin{align}
    h_{c}(f) = A_{\rm GW} \left( \frac{f}{f_{\rm yr}} \right)^{\alpha},
\end{align}
which is related to $\Omega_{\rm GW}$ as $\Omega_{\rm GW} = 2\pi^2 f^2 h_c^2 / (3H_0^2)$.
The relation between $\alpha$ and the slope of the cross-power spectral density $\gamma=3-2\alpha$
is also useful for direct comparison to the observational implication.
Since $\Omega_{\rm GW}(f) \propto f^{n_t}$ is observed
for the modes which re-enter the horizon in the radiation-dominated era,
we get $\alpha = (n_t-2)/2 = -{p}/{(p+1)}$ from Eq.~\eqref{eq:tensor-index},
if the extremely blue-tilted gravitational waves from the NEC-violating phase 
give rise to the spectrum around $f_{\rm yr}$.
In Fig.~\ref{fig:Energy_nt95}, we plot the
energy density spectra for different values of $p$.
We also show the sensitivity curves for present and future experiments 
of pulsar timing arrays and space/ground-based interferometers.
For the plotted spectra,
their spectral slopes and amplitudes are 
$(\gamma, \log_{10} A_{\rm GW})= (4.2,-14.525)$, 
$(4.1 , -14.542)$, and $(4.05 , -14.535)$, for $n_t=0.8$, $0.9$, and $0.95$, respectively, 
which are within the 2-$\sigma$ posterior contour 
for the five frequency power-law model 
of a spatially-uncorrelated common-spectrum process~\cite{Arzoumanian:2020vkk}.

Finally, let us discuss the BBN constraint~\cite{Allen:1996vm,Maggiore:1999vm,Smith:2006nka,Boyle:2007zx,Kuroyanagi:2014nba} in our model.
We calculate the total energy density of gravitational waves for the case with $n_t=0.8$ in Fig.~\ref{fig:Energy_nt95} 
and obtain its contribution to the effective number of relativistic species $\Delta N_{\rm eff}$ as
\begin{align}
     \Delta N_{\rm eff} \simeq \int \D
     \ln{\!f} \; \frac{\Omega_{\rm GW}h^2}{5.6 \times 10^{-6}} \simeq  1.3 \times 10^{-3}.
\end{align}
This is much smaller than the upper bound from the BBN constraint,
$\Delta N_{\rm eff} < \mathcal{O}(0.1)$~\cite{Cooke:2017cwo,Hsyu:2020uqb}.
(The similar bound
also comes from the large scale observations~\cite{Aghanim:2018eyx,Aiola:2020azj}.)
Since we have the nearly flat part of the spectrum in the high-frequency range,
we can avoid the problem pointed out by~\cite{Vagnozzi:2020gtf}, 
in which a single power-law, blue-tilted spectrum violates the BBN constraint
if it is consistent with both the CMB constraint and the NANOGrav result.

\section{Scalar Perturbations}\label{sec5}

Large-scale curvature perturbations must be generated
in the early universe to account for the observed nearly scale-invariant
spectrum of fluctuations. In our scenario, however, it cannot be due to
the fluctuations of the $\phi$ field, because in the NEC-violating phase
we have ${\cal G}_S=\,$const and ${\cal F}_S=\,$const, which means that
the curvature perturbation $\zeta$ has a highly blue spectrum
with $n_s-1=n_t>0$ and hence is irrelevant on CMB scales.

To generate nearly scale-invariant curvature perturbations on CMB scales,
we need another scalar field. Since the background in the NEC-violating phase
is significantly away from de Sitter,
a minimally coupled scalar field does not give rise to a scale-invariant spectrum
of fluctuations.
We therefore extend the mechanism proposed originally
in the context of the galilean genesis scenario~\cite{Creminelli:2010ba}
to a more general background, and
consider a non-minimally coupled scalar field,
\begin{align}
    S_\sigma=\int\D^4x\sqrt{-\widetilde g}\left[
    -\frac{1}{2}\widetilde g^{\mu\nu}\partial_\mu\sigma\partial_\nu\sigma-{\cal V}(\sigma) 
    \right],
\end{align}
with 
\begin{align}
    \widetilde g_{\mu\nu}=e^{2\phi/\mpl}g_{\mu\nu}.\label{confomal-metric}
\end{align}
The conformally related metric~\eqref{confomal-metric} reads more explicitly
\begin{align}
    \widetilde g_{\mu\nu}\D x^\mu\D x^\nu = 
    \widetilde a^2\left(-\D\eta^2+\D\Vec{x}^2\right),
\end{align}
where
\begin{align}
    \widetilde a =e^{\phi/\mpl}a \propto \frac{1}{(-\eta)},
\end{align}
in the NEC-violating phase, irrespective of the value of $p$.
This means that the non-minimally coupled scalar field $\sigma$
effectively lives in de Sitter, generating a nearly scale-invariant spectrum.
We therefore expect that fluctuations of the $\sigma$ field are the origin
of the observed spectrum of scalar perturbations.
To obtain a more detailed prediction, we need to investigate the
process of converting the fluctuations of $\sigma$ to adiabatic modes,
which is model-dependent and deserves further study.

\section{Conclusions}\label{sec6}

In this paper,
we have explored the possibility to generate nanohertz gravitational waves
consistent with both of the CMB and BBN constraints. 
We investigated the scenario which results in an extremely blue-tilted spectrum appropriate for 
the CMB constraint at $f \lesssim 10^{-15} \;{\rm Hz}$ and the NANOGrav result at $f \sim 1/{\rm yr}$ and a (nearly) flat spectrum for $f\gtrsim 1$/yr
to evade the BBN constraint.
To produce an extremely blue spectrum, 
we considered an NEC-violating phase in the early universe
undergoing the (super-)accelerated expansion with $a\propto (-t)^p$ ($p>0$),
in which the tensor spectral index $n_t>0$
is related to the power-law index $p$ as in Eq.~\eqref{eq:tensor-index}.
To obtain a flat spectrum in a high frequency range, 
we examined the scenario in which slow-roll inflation follows after the NEC-violating phase and ends with standard reheating,
as sketched out in the potential of Fig.~\ref{fig:potential.pdf}.

In Sec.~\ref{sec3},
we proposed a successful model in the cubic Horndeski theory.
First, we gave the minimal action~\eqref{Lag:NEC-V} which can cause a stable NEC-violating phase. 
We obtained the relation between the model parameters $(\alpha,\beta)$ and the solution parameters $(p,q)$ in Eq.~\eqref{Eq:abpq},
and the stability condition in terms of $(p,q)$ in Eqs.~\eqref{Stab:1} and \eqref{Stab:2}.
From these conditions, we found that a solution with $p \le 1$ is always unstable within our model, 
while stable solutions with $p>1$ are indeed possible.
We expect that a stable model with $p \le 1$ can be realized by adding other higher-order terms in the action,
but the analysis would be more complicated.
Next, we assume that the potential 
and the coefficient of the kinetic term are given by Eq.~\eqref{Func:kinatic}, 
which changes the field dynamics for $\phi > \phi_*$,
connecting the NEC-violating phase to the subsequent inflationary phase.
The phenomenological parameters $(H_\ast, H_{\rm inf}, \Delta{\cal N}_{\rm inf})$ are estimated in 
Eqs.~\eqref{Estimate:H-ast}, \eqref{Estimate:H-inf}, and \eqref{Estimate:N-inf},
enabling us to design the gravitational wave spectrum.
In Fig.~\ref{fig:time-evolution-of-vars},
we numerically gave the typical time evolution of the background
and showed that a gradient instability occurs in the intermediate decelerating phase.
We have discussed that this issue can be resolved by higher-order operators such as
in beyond-Horndeski theories~\cite{Creminelli:2016zwa,Cai:2016thi,Cai:2017tku,Kobayashi:2015gga}.

In Sec.~\ref{sec4},
we have presented the gravitational wave spectrum~\eqref{Def:GWspectrum} in our model with different parameter choices.
As is expected, those spectra are composed of
the blue-tilted part and the nearly flat part, 
of which amplitudes can be 
controlled almost independently of each other.
We gave gravitational wave spectra with $n_t = 0.8$, $0.9$, and $0.95$
for $f\lesssim 1/$yr
in Fig.~\ref{fig:Energy_nt95}, which are consistent with the recent NANOGrav result~\cite{Arzoumanian:2020vkk}.
We found that thanks to the flat part of the spectrum we are no longer bothered by the BBN bound and 
this flat part is possibly detected by future space/ground-based observations.
In Sec.~\ref{sec5}, we discussed how the nearly scale-invariant
scalar perturbations observed by CMB experiments are produced
by introducing another non-minimally coupled scalar field~\eqref{Lag:full}
which effectively lives in de Sitter even in the NEC-violating phase.

In summary, we have obtained an explicit early universe model
that can yield an extremely blue gravitational wave spectrum peaking
at $f\sim 1/$yr explored by pulsar timing arrays, while being consistent
with other observational constraints.

\acknowledgments
We thank Kohei Kamada and Sachiko Kuroyanagi for useful discussions.
HWHT was supported by the Grant-in-Aid for JSPS Research Fellow No.~JP19J11640.
TK was partially supported by JSPS KAKENHI Grant Nos.~JP20H04745 and~JP20K03936.

\appendix

\section{General background and perturbation equations %
for the cubic Horndeski theory}

We replicate the background and perturbation equations
for cubic Horndeski cosmology based on the Lagrangian~\eqref{cubic-Horndeski}.
The results are well known in the literature (see
Refs.~\cite{Kobayashi:2011nu,Deffayet:2010qz,Kobayashi:2010cm}).

The background equations for the metric $\D s^2=-\D t^2+a^2(t)\D\Vec{x}^2$
and the scalar field $\phi=\phi(t)$ are
given by
\begin{align}
    &3\mpl^2H^2 =2XG_{2X}-G_2+6HX\dot\phi G_{3X}-2XG_{3\phi},
    \label{app:Fridmann}
    \\
    &-\mpl^2\dot H =XG_{2X}+3HX\dot\phi G_{3X}-X\left(2G_{3\phi}+\ddot\phi G_{3X}\right),
    \label{app:Evolution}
    \\
    &\frac{1}{a^3}\frac{\D}{\D t}\left[
    a^3\left(\dot\phi G_{2X}+6HXG_{3X}-2\dot\phi G_{3\phi}\right)
    \right]
    \notag \\ &
    =G_{2\phi}-2X\left(G_{3\phi\phi}+\ddot\phi G_{3\phi X}\right).
    \label{app:Scalar}
\end{align}

Since $\phi$ is minimally coupled to the curvature tensor,
the equation for tensor modes, $h_{ij}(t,\Vec{x})$,
is the same as the standard one known in general relativity.
The quadratic action for the curvature perturbation in the
unitary gauge, $\zeta(t,\Vec{x})$, is given by
\begin{align}
    S_\zeta^{(2)}=\mpl^2\int\D t\D^3x\,a^3\left[
    {\cal G}_S\dot\zeta^2-\frac{{\cal F}_S}{a^2}(\partial\zeta)^2
    \right] ,
\end{align}
where 
\begin{align}
    {\cal G}_S&=\frac{1}{a}\frac{\D}{\D t}
    \left(\frac{a}{H-\dot\phi XG_{3X}/\mpl^2}\right)-1
    \\
    {\cal F}_S&=\frac{\Sigma}{(H-\dot\phi XG_{3X}/\mpl^2)^2},
\end{align}
with
\begin{align}
    \mpl^2\Sigma&=XG_{2X}+2X^2G_{2XX}+6H\dot\phi X G_{3X}
    \notag \\ & \quad 
    +6H\dot\phi X^2G_{3XX}-2XG_{3\phi}
    -2X^2G_{3\phi X}
    \notag \\ & \quad 
    +6X^3G_{3X}^2/\mpl^2.
\end{align}
The background under consideration is stable if
${\cal G}_S>0$ and $c_s^2:={\cal F}_S/{\cal G}_S>0$.

\bibliography{nanog}
\bibliographystyle{JHEP}
\end{document}